\documentclass[aps, prl, amsmath, amssymb, twocolumn, a4paper, superscriptaddress]{revtex4-2}
\usepackage{graphicx} 
\usepackage{dcolumn} 
\usepackage{bm} 
\usepackage{hyperref} 
\usepackage{color} 
\usepackage{multirow}
\makeatletter
\g@addto@macro\bfseries{\boldmath}
\makeatother

\def\pzo{Pr$_2$Zr$_2$O$_7$}

\begin{document}
\title{Intrinsic disorder in the candidate quantum spin ice \pzo}
\author{T.~J.~Hicken}
\affiliation{PSI Center for Neutron and Muon Sciences CNM, 5232 Villigen PSI, Switzerland}
\affiliation{Department of Physics, Royal Holloway University of London, Egham, TW20 0EX, United Kingdom}
\author{P.~Meadows}
\affiliation{Department of Physics, Royal Holloway University of London, Egham, TW20 0EX, United Kingdom}
\author{D.~Prabhakaran}
\affiliation{Clarendon Laboratory, University of Oxford Physics Department, Parks Road, Oxford, OX1 3PU, United Kingdom}
\author{A.~Szab\'o}
\affiliation{Max-Planck-Institut für Physik komplexer Systeme, Nöthnitzer Str.\ 38, 01187 Dresden, Germany}
\affiliation{Physik-Institut, Universität Zürich, Winterthurerstr.\ 190, 8057 Zürich, Switzerland}
\author{S.~E.~Dutton}
\affiliation{Cavendish Laboratory, University of Cambridge, Cambridge, CB3 0HE, United Kingdom}
\author{C.~Castelnovo}
\affiliation{TCM Group, Cavendish Laboratory, University of Cambridge, Cambridge, CB3 0HE, United Kingdom}
\author{K.~Moovendaran}
\altaffiliation[Current address: ]{Department of Physics and Nanotechnology, SRM Institute of Science and Technology, Kattankulathur, Tamil Nadu, India 603203}
\affiliation{Clarendon Laboratory, University of Oxford Physics Department, Parks Road, Oxford, OX1 3PU, United Kingdom}
\author{T.~S.~Northam de la Fuente}
\altaffiliation[Current address: ]{Materials Physics Center, CSIC-UPV/EHU, Paseo de Manuel Lardizabal 5, 20018 Donostia - San Sebastian, Spain}
\affiliation{Department of Physics, Royal Holloway University of London, Egham, TW20 0EX, United Kingdom}
\author{L.~Mangin-Thro}
\affiliation{Institut Laue-Langevin, 71 avenue des Martyrs, CS 20156, 38042 Grenoble C\'edex 9, France}
\author{G.~B.~G.~Stenning}
\affiliation{ISIS Facility, Rutherford Appleton Laboratory, Chilton, Didcot, OX11 0QX, United Kingdom}
\author{M.~J.~Gutmann}
\affiliation{ISIS Facility, Rutherford Appleton Laboratory, Chilton, Didcot, OX11 0QX, United Kingdom}
\author{G.~Sala}
\affiliation{Spallation Neutron Source, Second Target Station, Oak Ridge National Laboratory, Oak Ridge, Tennessee 37831, USA}
\author{M.~B.~Stone}
\affiliation{Neutron Scattering Division, Oak Ridge National Laboratory, Oak Ridge, Tennessee 37831, USA}
\author{P.~F.~Henry}
\affiliation{ISIS Facility, Rutherford Appleton Laboratory, Chilton, Didcot, OX11 0QX, United Kingdom}
\author{S.~A.~Barnett}
\affiliation{Diamond Light Source, Harwell Science and Innovation Campus, Chilton OX11 ORZ, United Kingdom}
\author{D.~R.~Allan}
\affiliation{Diamond Light Source, Harwell Science and Innovation Campus, Chilton OX11 ORZ, United Kingdom}
\author{D.~J.~Voneshen}
\affiliation{ISIS Facility, Rutherford Appleton Laboratory, Chilton, Didcot, OX11 0QX, United Kingdom}
\affiliation{Department of Physics, Royal Holloway University of London, Egham, TW20 0EX, United Kingdom}
\author{J.~P.~Goff}
\affiliation{Department of Physics, Royal Holloway University of London, Egham, TW20 0EX, United Kingdom}

\begin{abstract}
	\pzo\ is a promising example of a quantum spin liquid, where structural disorder plays a key role in enhancing quantum mechanical effects.
	Here we have determined the intrinsic defect structure of \pzo\ using neutron and x-ray scattering techniques supported by density functional theory.
	Our results explain the crystal field excitations by considering the non-magnetic singlets that arise on neighbouring sites.
	This makes a significant contribution towards the observed broadening of pinch points in the magnetic diffuse scattering, which was previously attributed purely to quantum effects.
\end{abstract}

\date{\today}
\maketitle

The search for materials that exhibit long-range quantum entanglement for applications in quantum technology has led to renewed interest in quantum spin liquids~\cite{savary2016quantum,zhou2017quantum,knolle2019field,broholm2020quantum}.
However, the delicate balance between competing interactions in geometrically frustrated magnets can be dramatically affected by the presence of structural disorder~\cite{ramirez2025short}, completely changing the responses of thermodynamic probes~\cite{chamorro2020chemistry} and even resulting in the stabilisation of long-range magnetic order~\cite{zheng2006coexisting,koushik2025emergence}.
Structural disorder is often regarded as a nuisance~\cite{broholm2020quantum,syzranov2022eminuscent} since it is believed to lead to decoherence, and much effort has been devoted to eliminating it.
Recently it has been proposed that for materials with non-Kramers ions, strains can introduce splitting of the ground state doublet akin to the effect of a local transverse field, so that structural disorder actually induces quantum spin liquid behaviour~\cite{savary2017disorder,benton2018instabilities}.

In the case of pyrochlores, the importance of understanding defect structures is becoming increasingly recognised.
For the classical spin ice material Dy$_2$Ti$_2$O$_7$ which exhibits emergent, fractionalised, magnetic monopole excitations, relatively low levels of oxygen vacancies are found to control the magnetic monopole dynamics~\cite{sala2014vacancy}.
Minute levels of intersite mixing of cations in Tb$_2$Ti$_2$O$_7$ tunes the sample between an ordered and a disordered phase~\cite{taniguchi2013long}, and in Yb$_2$Ti$_2$O$_7$ oxygen vacancies change the ground state from ferromagnetic to a spin liquid~\cite{bowman2019role}.
Some of the most suggestive evidence to date of quantum spin liquid behaviour induced by structural disorder has been provided by studies of \pzo~\cite{kimura2013quantum} where the presence of intrinsic structural disorder comes into play~\cite{wen2017disordered,martin2017disorder,trump2018universal}.

One key probe of frustrated materials is magnetic diffuse scattering, which is sensitive to short-range magnetic correlations.
In \pzo, most of the magnetic diffuse scattering is inelastic and, in contrast to a classical spin ice, the pinch points are broadened, giving a starfish-like pattern~\cite{kimura2013quantum}.
The inelastic neutron scattering from \pzo\ has been modelled in terms of a random transverse field distribution presumed to result from structural disorder~\cite{wen2017disordered}, as well as the random strain distribution itself determined by structural diffuse neutron scattering~\cite{martin2017disorder}.
Gaining a full understanding of the defect structures, and their experimental signatures, is therefore of the utmost importance.

The presence of relatively low levels of structural disorder typically leads to small changes in Bragg intensities, making it challenging to determine defect structures using standard diffraction techniques.
In contrast, the structural diffuse scattering between Bragg reflections arises from structural disorder and is highly sensitive to the defect structure.
Nevertheless, the need to include small displacements of many ions surrounding the defects via Monte Carlo modelling~\cite{sala2014vacancy} can make it challenging to calculate and interpret diffuse scattering.
Previous measurements of structural diffuse scattering from \pzo\ were analysed in terms of the Huang scattering near Bragg reflections, which arises from elastic displacement fields around defects~\cite{martin2017disorder}.
However, Huang scattering is inversion symmetric~\cite{dederichs1973theory} whereas the reported scattering does not show evidence of such symmetry, suggesting that the defect structure is more complicated.
Furthermore, the diffuse scattering is mainly away from Bragg peaks (both in Ref.~\cite{martin2017disorder} and in our measurements detailed below), which arises primarily from displacements of ions in the vicinity of defects.
As such, the origin of the diffuse scattering, and how this corresponds to the defect structures in \pzo, remains unclear.

In this Letter we determine the intrinsic defect structures in \pzo\ using ab initio density functional theory (DFT).
Our results explain all the available diffraction, diffuse, and inelastic scattering data from \pzo, as well as bulk thermodynamic measurements.
Further, we demonstrate that broadening of the pinch points in the magnetic diffuse scattering is a natural consequence of the disorder.
Our results show that the magnetism in \pzo\ cannot be fully understood without taking account of the large numbers of robust nonmagnetic singlets generated by relatively low levels of structural disorder.

\begin{figure}
	\centering
	\includegraphics[width=\linewidth]{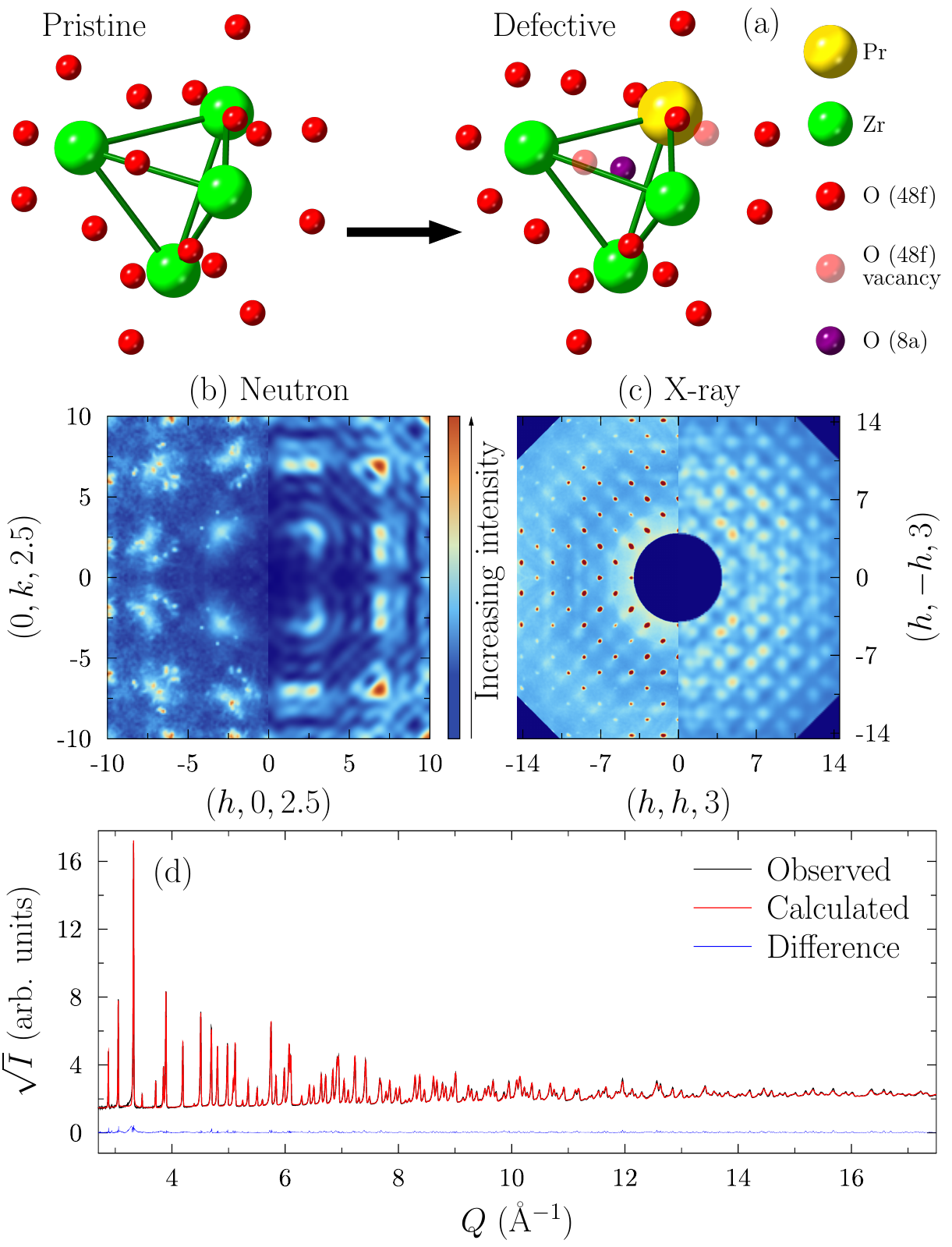}
	\caption{The fundamental defect unit of the intrinsic defect structure of \pzo, calculated with \textit{ab-initio} methods, is shown in (a, right), with atoms that are missing compared to the pristine crystal structure (a, left) shown as translucent. The defect structure consists of two Pr substitutions, one of which has an accompanying charge-compensating O 48f vacancy, with another O 48f ion relaxing towards a nearby O 8a site. The position of the nearby O ions also relax, breaking the local symmetry. Diffuse scattering patterns from \pzo\ in the $(h,k,n)$ planes are shown for both(b) neutron, $n=2.5$, and (c) x-ray, $n=3$, scattering. Comparisons to our calculations (right half of each plot) show excellent agreement (note that the experimentally observed Bragg peaks are intentionally omitted from our calculations). The average structure is understood through room temperature powder scattering, such as the neutron scattering shown in (d).}
	\label{fig:structure}
\end{figure}

We have explored potential intrinsic defect structures of \pzo\ from first-principles using DFT methods~\cite{sm} implemented in the \textsc{castep} code~\cite{clark2005first}.
There are many possible defect structures, constrained by the requirement of a charge neutral crystal; this means that Pr/Zr substitutions must be compensated by vacant or additional O ions.
For each of our trial defect structures, we identify the lowest energy configuration which is most likely to be experimentally realised.
We find that a single defect structure, shown in Fig.~\ref{fig:structure}(a), can well explain all the experimental measurements of \pzo.
This defect structure has one Pr atom on a Zr site, with two of the normally occupied O 48f sites unoccupied; between these two vacancies, an O interstitial sits close to the face of the Zr tetrahedra, near the 8a Wyckoff position.
Similar O distortions have been reported in other pyrochlore materials and are often associated with cation disorder~\cite{heremans1995fast,west2008structural,mandal2010order,blanchard2013investigating,miruszewski2024high,fuentes2024pyrochlore}.
Other nearby O ions also have smaller, but noticeable, changes in their positions.
Another Zr to Pr substitution occurs elsewhere in the unit cell, but appears to be uncorrelated to the main defect structure and have little impact on the surrounding ions.

Deviations from a perfect crystal symmetry primarily contribute to diffuse scattering, hence we have performed such measurements on single crystal samples produced using the floating-zone technique~\cite{sm} using both lab-based x-ray and facility-based neutron synchrotron instruments (the SXD instrument~\cite{keen2006sxd} at the ISIS Neutron and Muon Source, UK).
The features in the measured patterns are highly sensitive to the defect structures and their associated distortions, and whilst the existence of diffuse scattering unambiguously demonstrates the existence of disorder in \pzo, it is generally hard to gain quantitative information from these measurements.
To tackle this problem, we have calculated the diffuse scattering for a \pzo\ crystal containing the DFT-calculated defect structure, including the ion displacements found from the geometry optimisation.
The scattering from the defect structure in Fig.~\ref{fig:structure}(a) well describes the experimental neutron and x-ray scattering in all planes~\cite{sm}; we show one such plane for each probe in Fig.~\ref{fig:structure}(b--c).
The lack of superstructure peaks and sharp features in the experimental diffuse scattering allow us to confidently rule out the possibility of a more complex, correlated disorder~\cite{goff1999defect}.
No other trial defect structure explains the diffuse scattering as well~\cite{sm}.
Whilst reverse Monte Carlo methods very effectively model diffuse scattering patterns~\cite{nield1995interpretation,proffen1997discus,morgan2021rmc}, compared to other approaches, this first-principles method provides greater physical insight into the realised defect structure which can inform other measurements and calculations.

X-ray diffraction is the standard structural characterisation technique, and both powder and single-crystal studies have been reported for \pzo~\cite{kimura2013quantum,hatnean2014structural,koohpayeh2014synthesis}.
These studies show a low level ($\simeq3$\%) of intersite mixing of Pr and Zr ions that varies with sample, but they are not sensitive to O stoichiometry.
Our powder x-ray diffraction studies of \pzo\ confirm the absence of impurity phases, and are well described with an ideal stoichiometry fit~\cite{sm} and isotropic thermal parameters.
Our single-crystal x-ray diffraction data are also fitted with ideal stoichiometry, but anisotropic thermal parameters are essential to account for the data.
We find that the thermal ellipsoids for O 48f ions are elongated in the direction of the neighbouring 8a site~\cite{sm}.
This elongation is in agreement with our DFT geometry optimisation, where we find the nearest O 48f ions are shifted by an average of approximately 0.25~\AA\ away from the now-occupied O 8a site. 

The greater sensitivity of neutrons to the presence of O vacancies and interstitials, combined with the much wider $Q$-range of our powder neutron diffraction data, better constrains the defect model.
The fit to such data (measured with the Polaris diffractometer at the ISIS Neutron and Muon Source, UK~\cite{smith2019upgraded,sm}) for one detector bank is shown in Fig.~\ref{fig:structure}(d) with the fits to the whole dataset and details of the refinement in Ref.~\cite{sm}.
We find occupation of the Zr site by excess Pr ions, with accompanying vacancies on the O 48f site, and O interstitials on the 8a site.
The relative occupancies are consistent with the defect model determined by DFT [Fig.~\ref{fig:structure}(a)], with 5.5(6)\% of the tetrahedra having this defect structure.
In the following sections we demonstrate that the effects of this relatively small deviation from the perfect structure are far more significant than one might anticipate.

\begin{figure}
	\centering
	\includegraphics[width=\linewidth]{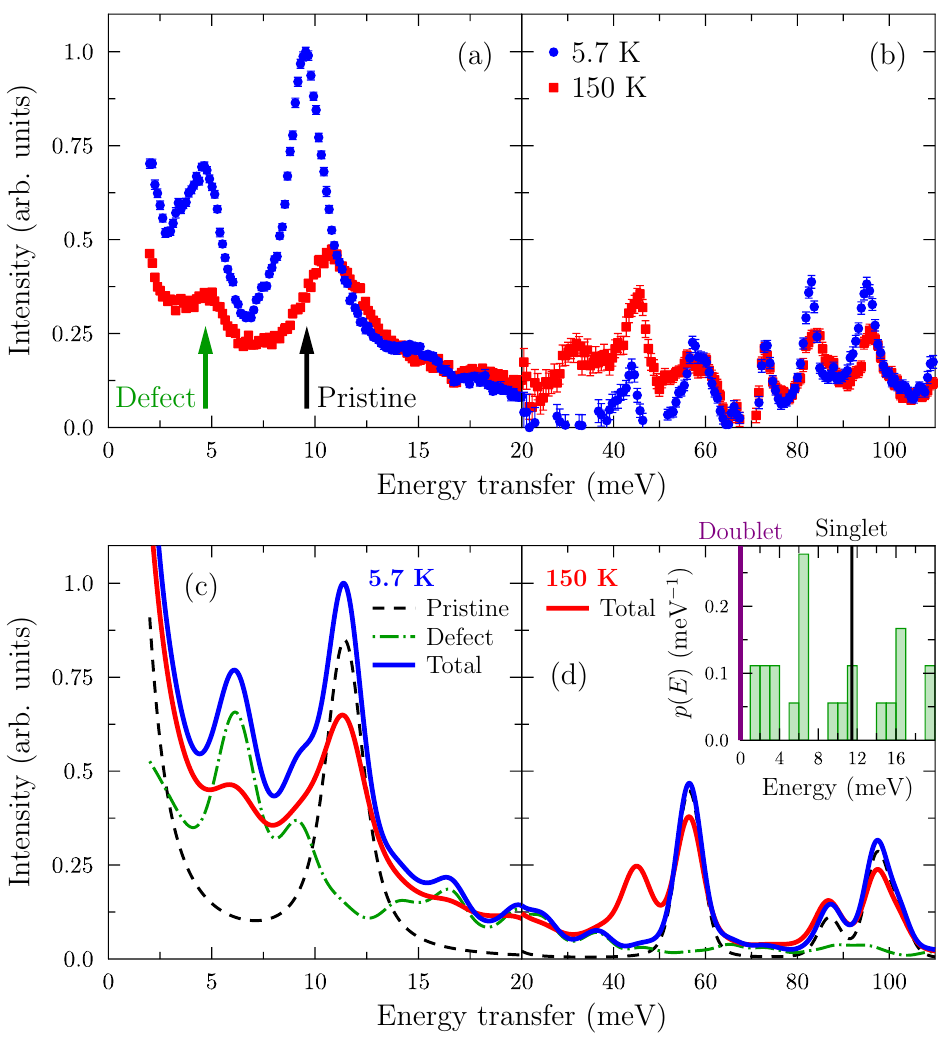}
	\caption{The crystal electric field energy levels of \pzo\ can be directly probed (a--b) by inelastic neutron scattering, and compared to our point charge model predictions (c--d), where the effects of both a pristine and defective local structure are shown individually for the low-temperature simulation. The lowest energy peak (a/c, green arrow), predominantly arises due to the defect structure, whereas the second peak (black arrow), predominantly arises from the pristine structure. Note the change in the horizontal axis scale to show data with different incident neutron energies. The inset in (d) shows the probability density of finding an energy level gap between the ground state and an excited state in the defective unit cell, from the same point charge model. The peaks here correspond to the peaks seen in the inelastic neutron scattering in (c). The ground state doublet and first excited singlet in the pristine cell are indicated.}
	\label{fig:cef}
\end{figure}

Pr$^{3+}$ is a non-Kramers ion, meaning that it can exhibit either a singlet (non-magnetic) or doublet (magnetic) ground state, depending on the crystal symmetry.
In pristine \pzo\ a doublet ground state is expected~\cite{kimura2013quantum,bonville2016magnetic}, and this is key for the interesting magnetic properties of the material.
However, when the intrinsic defect structures are present, Pr environments near these defects have a lowered symmetry.
Therefore, even at a low defect level, we find that a large fraction of the Pr ions are substantially affected.
In a back-of-the-envelope estimate one would expect around 30\% of Pr ions to be affected, even without considering further neighbour effects.
For the defect cluster in Fig.~\ref{fig:structure}(a), each O 8a interstitial has 12 nearest-neighbour Pr ions, which include the 4 Pr ions that are nearest neighbours to the O 48f vacancies.
Hence, since there is one such defect cluster for every two stuffed ions, 5\% stuffing implies 30\% of Pr ions are nearest neighbours to such a defect cluster.

To understand these intrinsic defect-induced changes in the single-ion magnetism quantitatively, we have performed point-charge model calculations~\cite{sm} that give the ground state and crystal-electric field excitations for each Pr atom.
We find that close to the defect structure the Pr atoms exhibit singlet ground states, with a distribution of gaps to the first excited state ranging between $\simeq1.2$~meV and $\simeq20$~meV, with an average gap size of around $7.5$~meV.
This distribution of excitations reveals that energy levels in the defect structure are different to, but clustered around, those of the pristine structure.
A distribution of energies between the ground and first excited state has been previously suggested~\cite{wen2017disordered}, however our results show that the gaps are much larger than the previously suggested average of 0.27(1)~meV.
Crucially, the energy of the first excited state is significantly larger than the exchange, dipolar, or other interactions in the system, and hence the single ion magnetism of these lower symmetry sites has a spin singlet character.

To verify that these calculations match experimental observations, we have performed inelastic neutron scattering (INS) measurements~\cite{sm} of powder \pzo\ using the SEQUOIA instrument~\cite{granroth2010sequoia} at the Spallation Neutron Source, Oak Ridge National Laboratory, USA.
In these measurements, the energy transferred from the incoming neutron to the sample excites electrons to higher crystal-field energy levels, hence by measuring the transferred energy we can directly probe the crystal-electric field energy levels.
Such measurements can be seen in Fig.~\ref{fig:cef}(a--b), with comparison to our calculations from the point-charge model of the pristine and defect structures.
It is clear that the defect structure can explain the low energy peak around 4~meV (which is below the minimum energy transfer measured in other work, such as Ref.~\cite{kimura2013quantum}); the peak represents the average energy of the first excited state when the Pr has a singlet ground state.
The two weak peaks between 60 and 75~meV, also observed in Ref.~\cite{kimura2013quantum}, can equally be explained by the inclusion of the defect structure (although the peak resolution in the simulations prevents the peaks from being observed easily).
In fact, every magnetic peak in the observed spectrum now well corresponds to a peak in our calculated spectrum.
Further, the temperature dependence of the INS spectrum, which arises due to changing thermal populations of the crystal electric field energy levels, is also well captured by our model.

With the energy levels correctly captured by our defect model, comparing the intensity of these peaks to experimental measurements provides an estimate the fraction of Pr that are affected and possess a singlet ground state.
By comparing the 4~meV peak (arising from singlet Pr sites) and the 10~meV peak (arising from doublet Pr sites), we find that approximately 30\% of the Pr sites in the material are singlets (specifically, these are sites where the gap between the ground and first excited state is significantly larger than the exchange, dipolar, or other interactions, hence they cannot be considered as quasi-doublets.)
We note that 30\% matches the intuitive suggestion made earlier.
This demonstrates that the defect structure (which affects around 6\% of the tetrahedra) has significant effects on the magnetism, with a much higher fraction of Pr sites best described as singlets.
This ratio of singlet to doublet sites gives reasonable peak intensities over the entire energy range, see Fig.~\ref{fig:cef}(c--d).
Interestingly, Ref.~\cite{kimura2013quantum} introduces a 37\% fraction of Pr sites with a reduced magnetic moment to explain the magnitude of their heat capacity data; the similarity in this number to our defective fraction suggests that our defect structure may be the physical interpretation of their findings.
Whilst the defect concentration will be sample dependent, it seems likely that this type of defect will be found in most \pzo\ samples.
As so many Pr ions are affected by such a small fraction of defects, the effects arising due to these defects are likely to be important in the majority of cases.

A key feature in the magnetic behaviour of \pzo\ is the spin-ice behaviour at low temperatures, which arises as a consequence of magnetic interactions between Pr ions that have doublet (magnetic) ground states.
Having demonstrated that a sizeable fraction of the Pr sites are singlets, we must consider the consequences this will have on the collective magnetic behaviour.
We start by probing the system experimentally to verify that our samples observe the same magnetic response as previously reported.
Our AC susceptibility measurements below 2~K (performed at the ISIS Neutron and Muon Source Materials Characterisation Laboratory, UK) identify a frequency-dependent peak with no signature of long range ordering~\cite{sm}, consistent with other observations~\cite{kimura2013quantum} and explained as thermal activation often seen in spin ice materials.
The Curie-Weiss temperature (fitted below 10~K to minimise the impact of crystal electric field effects) is approximately 1.5~K, within the reported range~\cite{matsuhira2009spin,kimura2013quantum,hatnean2014structural,petit2016antiferroquadrupolar,wen2017disordered,tokiwa2018discovery}.
These measurements, alongside the now-explained clear features of the defect structure seen in other works, make us confident that our sample has the same magnetic response as those in other works.

\begin{figure}
	\centering
	\includegraphics[width=\linewidth]{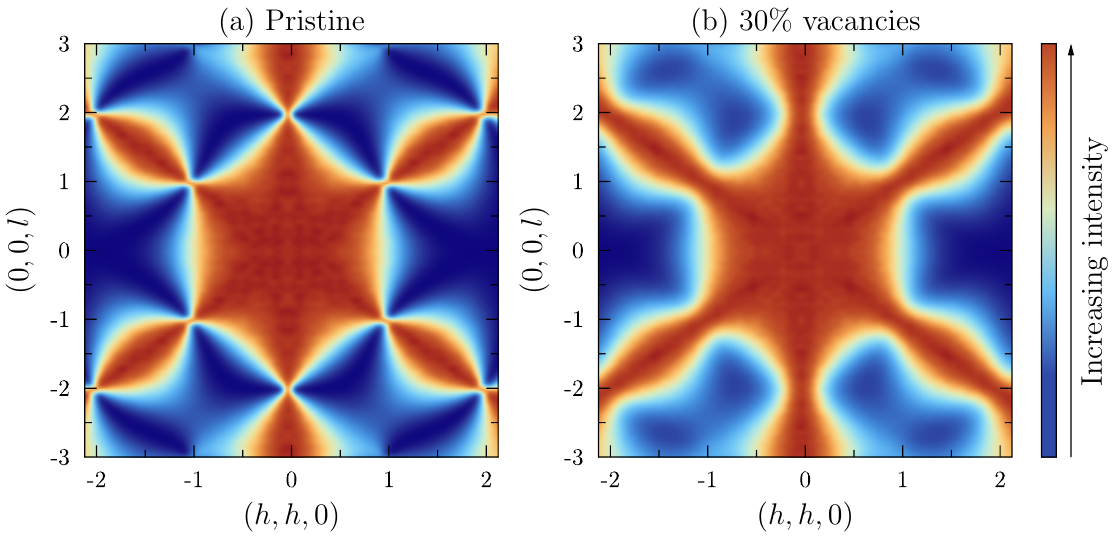}
	\caption{Simulations of the magnetic diffuse scattering calculated (left) from simulations of the magnetic state in a pristine pyrochlore (i.e., one with no defects) at $T=0.1$J, equivalent to approximately 100~mK in our system. Similar simulations were performed with 30\% of the magnetic moments randomly removed (right). The broadening of the pinch points is reminiscent of the experimentally observed patterns in Refs.~\cite{kimura2013quantum,petit2016antiferroquadrupolar}.}
	\label{fig:collective}
\end{figure}

We model the effect of the Pr singlets using classical Monte Carlo simulations~\cite{sm}, similar to those that have been performed for the classical spin ice materials Dy$_2$Ti$_2$O$_7$ and Ho$_2$Ti$_2$O$_7$~\cite{den2000dipolar,bramwell2001spin,jaubert2009signature}.
In these calculations, we approximate the spin coupling using a nearest-neighbour Ising interaction between the low-energy doublets, with the defect structure taken into account by removing sites where the ground state doublet is split.
These simulations are used to sample the system configurations in equilibrium at the desired temperature, and to compute the magnetic diffuse scattering patterns measured experimentally.
Upon introducing non-magnetic sites in the pyrochlore lattice we find that the pinch points observed in the spin-flip channel of the magnetic diffuse scattering pattern are broadened, as shown in Fig.~\ref{fig:collective}.
We find similar broadening of the pinch points for other physically reasonable models of magnetic vacancies.
Broadening of the pinch points is taken as the signature of quantum fluctuations in \pzo~\cite{kimura2013quantum,petit2016antiferroquadrupolar,wen2017disordered,martin2017disorder}, and is the most important experimental finding that identifies the material as host to a quantum spin ice phase.
Our work demonstrates that the intrinsic defect structure of \pzo\ will naturally lead to broadened pinch points without the need to invoke quantum fluctuations.
Whilst this does not exclude the possibility of a quantum spin ice state in \pzo, taking this intrinsic broadening into account is essential to understand the state truly realised in the system, and to explain other features previously attributed to quantum fluctuations such as the excitation at 0.3~meV~\cite{kimura2013quantum,wen2017disordered,martin2017disorder}.
Further, as the broadening is likely robust to the precise details of the structural defects, it may be applicable to other non-Kramers systems.

Combining DFT and diffuse scattering is an approach that could be extended to other systems, for example other pyrochlores which are also quantum spin ice candidates.
Tb$_2$Hf$_2$O$_7$ exhibits a high level of Frenkel disorder (vacant O 48f sites with charge compensating O 8a interstitials), and it also has a defective crystal electric field environment for a large fraction of magnetic ions~\cite{sibille2017coulomb}.
Structural disorder was also reported for Pr$_2$Sn$_2$O$_7$, however it has not yet been possible to unambiguously determine the defect structure~\cite{luo2025disorder}.
A candidate for a quantum spin ice with minimal defects is Pr$_2$Hf$_2$O$_7$, which has been reported to have a very low level of structural disorder~\cite{sibille2018experimental}.
Our own measurements~\cite{sm} reveal extensive diffuse scattering; however we identify that this is dominated by phonon scattering.
Thus, the level of intrinsic structural disorder in Pr$_2$Hf$_2$O$_7$ is likely low, and this material may serve as an excellent platform to explore quantum spin ice physics.
In all pyrochlore systems, it is clear that high-purity, defect-free systems are needed to study quantum effects.
Beyond pyrochlores, YbMgGaO$_4$ has chemical disorder that mimics the experimental signatures of a spin liquid~\cite{zhu2017disorder}.
Indeed, many of the most common quantum spin liquid candidate materials (such as herbertsmithite~\cite{freedman2010site,nilsen2013low,zorko2017symmetry}, barlowite~\cite{liu2015selectively,smaha2018synthesis}, and FeSc$_2$S$_4$~\cite{tsurkan2017structure}) can show a high degree of structural disorder~\cite{chamorro2020chemistry}, and may warrant further exploration with the techniques developed here.

Our work has demonstrated the high sensitivity of the \pzo\ system to intrinsic disorder.
The profound impact that we have uncovered is quite surprising, especially given the deviation from a perfect lattice is very small, no more than a few percent.
Most significantly, our work suggests that the broadening of the pinch points in the magnetic diffuse scattering can arise due to intrinsic disorder, and is not necessarily the ``smoking-gun'' evidence of a quantum spin ice it was previously believed to be.
Our results also demonstrate the importance of carefully exploring the crystal structure of materials that may host exotic magnetic states, as these deviations from a perfect crystal are easily missed by conventional experimental probes.
Rather than view these imperfections as a nuisance, we propose that this presents an opportunity to tune the magnetism in \pzo\ and similar systems~\cite{savary2017disorder}, for example through chemical substitution, allowing us to access a far greater number of magnetic states than would be possible if constrained to perfect crystal structures.
These states may have technological benefits~\cite{gliga2020dynamics}, but also allow us to test our fundamental understanding of the physics behind these magnetic systems by tailor-making samples that can manifest a particular theoretical model.

The insights in this paper were enabled by the utilisation of DFT approaches to simulate the results of experimental measurements.
Specifically, whilst DFT calculations typically allow one to assess the stability of defect structures by looking at the total energy, we have demonstrated that, by predicting the diffuse scattering directly from the geometry optimised structure, one can additionally compare it with the experimentally measured structural relaxations.
Understanding the role structural defects play in two-level systems is of key importance in limiting decoherence in quantum sensors.
For the wider area of frustrated magnetism, the relevance of structural disorder inferred from the sample dependence of measurements, the need is to determine the details of the defect structure and, from this, to simulate the physical properties.
Lattice imperfections are important in many other areas of physical science, and it affects physical properties such as diffusion pathways in sensors, fuel cells and battery materials, material plasticity and strength, and device fabrication.
The use of DFT significantly simplifies the analysis of diffuse scattering data to the point where this uniquely sensitive probe of defect structure can be applied to a wide range of research fields and technologies.

\section*{Acknowledgements}
The Engineering and Physical Sciences Research Council (EPSRC) supported the work performed at Royal Holloway, University of London (EP/T028041/1, EP/X034526/1), the University of Oxford (EP/T028637/1), and the University of Cambridge (EP/T028580/1).
DP acknowledges the Oxford-ShanghaiTech collaboration project for financial support.
AS was supported by Ambizione grant No.~215979 by the Swiss National Science Foundation.
We are grateful to all facilities that have provided time for experiments or calculations in this work, including; beam time on the SXD and Polaris instruments~\cite{dois} at the ISIS Neutron and Muon Source, UK, and further measurements at the Materials Characterisation Laboratory; computing resources from the UK STFC Scientific Computing Department’s SCARF cluster; the Spallation Neutron Source, a DOE Office of Science User Facility operated by the Oak Ridge National Laboratory, USA, where the beam time was allocated to SEQUOIA on proposal number IPTS-27001.1; and on I19 at the Diamond Light Source under proposal CY44441.
We would like to thank Li Ern Chern for useful discussions during this project.

\section*{Author contributions}
TJH conceived and performed the density functional theory calculations.
DP and KM synthesised the samples.
PM, MJG, and JPG performed the diffuse scattering measurements which were subsequently analysed by PM.
TJH calculated the theoretical diffuse scattering patterns.
PM, TJH, and JPG performed and analysed the powder x-ray diffraction measurements.
PM, DJV, and JPG performed and analysed the lab-based single-crystal x-ray diffraction experiment.
SAB, DRA, PM, and JPG performed and analysed the synchrotron experiment.
TJH, PM, PFH, SED, and JPG performed and analysed the powder neutron diffraction measurements.
TJH, PM, DP, and GBGS performed the magnetometry measurements which were subsequently analysed by TJH.
GS, MBS, TSNdlF, LMT, and JPG performed the inelastic neutron scattering measurements which were subsequently analysed by TJH and DJV.
TJH performed the point charge model calculations.
AS and CC performed the theoretical magnetic diffuse scattering calculations.
JPG, CC, SED, and DP conceived the project, which was then developed by TJH and JPG.
All authors discussed the results.
The manuscript was originally written by TJH and JPG, and all authors contributed to the revisions.

\bibliography{bib}

\end{document}


\title{Supplemental Material for ``Intrinsic disorder in the candidate quantum spin ice \pzo''}
\author{T.~J.~Hicken}
\affiliation{PSI Center for Neutron and Muon Sciences CNM, 5232 Villigen PSI, Switzerland}
\affiliation{Department of Physics, Royal Holloway University of London, Egham, TW20 0EX, United Kingdom}
\author{P.~Meadows}
\affiliation{Department of Physics, Royal Holloway University of London, Egham, TW20 0EX, United Kingdom}
\author{D.~Prabhakaran}
\affiliation{Clarendon Laboratory, University of Oxford Physics Department, Parks Road, Oxford, OX1 3PU, United Kingdom}
\author{A.~Szab\'o}
\affiliation{Max-Planck-Institut für Physik komplexer Systeme, Nöthnitzer Str.\ 38, 01187 Dresden, Germany}
\affiliation{Physik-Institut, Universität Zürich, Winterthurerstr.\ 190, 8057 Zürich, Switzerland}
\author{S.~E.~Dutton}
\affiliation{Cavendish Laboratory, University of Cambridge, Cambridge, CB3 0HE, United Kingdom}
\author{C.~Castelnovo}
\affiliation{TCM Group, Cavendish Laboratory, University of Cambridge, Cambridge, CB3 0HE, United Kingdom}
\author{K.~Moovendaran}
\altaffiliation[Current address: ]{Department of Physics and Nanotechnology, SRM Institute of Science and Technology, Kattankulathur, Tamil Nadu, India 603203}
\affiliation{Clarendon Laboratory, University of Oxford Physics Department, Parks Road, Oxford, OX1 3PU, United Kingdom}
\author{T.~S.~Northam de la Fuente}
\altaffiliation[Current address: ]{Materials Physics Center, CSIC-UPV/EHU, Paseo de Manuel Lardizabal 5, 20018 Donostia - San Sebastian, Spain}
\affiliation{Department of Physics, Royal Holloway University of London, Egham, TW20 0EX, United Kingdom}
\author{L.~Mangin-Thro}
\affiliation{Institut Laue-Langevin, 71 avenue des Martyrs, CS 20156, 38042 Grenoble C\'edex 9, France}
\author{G.~B.~G.~Stenning}
\affiliation{ISIS Facility, Rutherford Appleton Laboratory, Chilton, Didcot, OX11 0QX, United Kingdom}
\author{M.~J.~Gutmann}
\affiliation{ISIS Facility, Rutherford Appleton Laboratory, Chilton, Didcot, OX11 0QX, United Kingdom}
\author{G.~Sala}
\affiliation{Spallation Neutron Source, Second Target Station, Oak Ridge National Laboratory, Oak Ridge, Tennessee 37831, USA}
\author{M.~B.~Stone}
\affiliation{Neutron Scattering Division, Oak Ridge National Laboratory, Oak Ridge, Tennessee 37831, USA}
\author{P.~F.~Henry}
\affiliation{ISIS Facility, Rutherford Appleton Laboratory, Chilton, Didcot, OX11 0QX, United Kingdom}
\author{S.~A.~Barnett}
\affiliation{Diamond Light Source, Harwell Science and Innovation Campus, Chilton OX11 ORZ, United Kingdom}
\author{D.~R.~Allan}
\affiliation{Diamond Light Source, Harwell Science and Innovation Campus, Chilton OX11 ORZ, United Kingdom}
\author{D.~J.~Voneshen}
\affiliation{ISIS Facility, Rutherford Appleton Laboratory, Chilton, Didcot, OX11 0QX, United Kingdom}
\affiliation{Department of Physics, Royal Holloway University of London, Egham, TW20 0EX, United Kingdom}
\author{J.~P.~Goff}
\affiliation{Department of Physics, Royal Holloway University of London, Egham, TW20 0EX, United Kingdom}

\begin{abstract}
	We present detailed experimental and computational methods for all the techniques employed in our work, as well as additional diffuse scattering measurements and calculations, further details on our scattering experiments and refinements, measurements of the DC and AC susceptibilities of our \pzo\ sample, simulations of the contribution to the heat capacity arising from crystal field effects from the defect structure, and further discussion of the Monte Carlo calculations. In addition, we present diffuse scattering measurements of another quantum spin ice candidate material, Pr$_2$Hf$_2$O$_7$, which we demonstrate to be dominated by phonon scattering, rather than defects as in the case of \pzo.
\end{abstract}

\date{\today}
\maketitle

\section{Methods}
\subsection{Density functional theory}
Density functional theory (DFT) calculations were performed using the plane-wave pseudopotential code \textsc{castep}~\cite{clark2005first}.
The generalised gradient approximation (PBE)~\cite{perdew1996generalized} was used in all calculations.
A cell the size of the conventional unit cell ($\sim10.7\times10.7\times10.7$~\AA) was used.
Calculations were converged to better than 0.6~meV/atom using a plane-wave cutoff of 900 eV, $\vect{G}$-vectors up to 36~\AA$^{-1}$, and a $3\times3\times3$~$k$-point grid~\cite{monkhorst1976special}.
47 calculations were performed such that all symmetrically inequivalent unit cells of the same configuration were tried.
The atomic positions were subsequently allowed to relax whilst keeping the lattice parameter fixed to the value found for the pristine cell.
In a real system, the defect structures are expected to be randomly positioned and oriented, hence they will not introduce a strain in any special direction.
By keeping the lattice parameters fixed in our calculations, we ensure that there are no strain artefacts when only considering a single defect.
This relaxation stops once the energy and atomic positions have converged to better than $1\times10^{-6}$~meV/atom and $1\times10^{-4}$~\AA\ respectively, and the maximum force on any atom is less than $2\times10^{-3}$~eV/\AA.

\subsection{Crystal growth}
Powder samples of \pzo\ were synthesised using the traditional solid-state method by intimately mixing stoichiometric amounts of high purity  Pr$_6$O$_{11}$ (99.99\%) and ZrO (99.99\%).
Prior to weighing, the chemicals were dried overnight at 1000$^\circ$C to remove any moisture.
The sintering was performed in the form of pellets at 1400$^\circ$C for 24 hours with intermediate grinding; this procedure was repeated three times.
Finally, polycrystalline powder was pressed into cylindrical rods (6~mm diameter, 75~mm long) and sintered at 1450$^\circ$C in a high purity argon atmosphere for 24 hours.
This step is important to remove any Pr$^{4+}$ present in the sample; during this step, the colour of the powder changes from brown to light green.

Single crystals of \pzo\ were grown by the optical floating-zone technique using a 5~kW xenon arc furnace (HKZ, SciDre GmbH) under 10 atmosphere pure argon pressure with a growth rate of 12--15~mm/h.
During growth, afterheat was employed to reduce the cracking of the crystal.
There was no PrO$_2$ evaporation observed during growth.
The resulting crystals were found to be highly pure, possessing a light green colour consistent with previous reports.

\subsection{Diffuse scattering}
The diffuse neutron scattering from a crystal of \pzo\ was measured using the SXD instrument~\cite{keen2006sxd} at the ISIS Neutron and Muon Source, UK.
SXD combines the white beam Laue technique with area detectors covering a solid-angle of $2\pi$ steradians, allowing comprehensive data sets to be collected.
Samples were mounted on aluminium pins and cooled using a closed-cycle helium refrigerator.
Six orientations were collected for four hours per orientation.
Data were corrected for incident flux using a null-scattering V/Nb sphere.
These data were then combined to a volume in reciprocal space and sliced to obtain individual planar cuts.
Measurements were performed at 30 and 300~K, and the fact that the diffuse intensity was independent of temperature confirmed its origin as structural disorder rather than inelastic scattering.
We predominantly show room temperature data in this manuscript.
The scattering in planes of the form (h, k, $n$), where $n=[0,10]$, were analysed using the SXD2001 program~\cite{gutmann2005sxd2001}.
The voxel size was approximately $0.1$~\AA$^{-1}$ per axis.
All data was averaged according to Laue symmetry.

Single-crystal x-ray scattering measurements were performed on \pzo\ with x-rays at the Mo K edge at room temperature using an in-house Rigaku Xcalibur diffractometer.
Sub-millimetre single crystals were cleaved from the boule, and attached to glass capillaries using silicone grease.
Large volumes of reciprocal space were surveyed with a CCD (charge-coupled device) detector.
A very small sample was used for single-crystal x-ray diffraction in order to minimise absorption and extinction corrections.
These same measurements were used to refine the structure and obtain the anisotropic thermal parameters using the Jana2006 software~\cite{petvrivcek2014crystallographic}.
A larger crystal was used for single-crystal diffuse scattering in order to maximise the signal in comparison to the background from the sample mount.
The diffuse scattering data was converted to reciprocal space using the CrysAlisPro software, with a voxel size of approximately $0.01$~\AA$^{-1}$ per axis.
The data were averaged according to Laue symmetry.

Single-crystal x-ray scattering measurements were performed on Pr$_2$Hf$_2$O$_7$ using the I19 Small Molecule Single Crystal Diffractometer at the Diamond Light Source~\cite{allan2017novel}.
The energy of the x-ray beam was 17.9976~keV (the K edge of Zr), and a sphere of data was collected to give 0.6 Å resolution.
The measurements were performed in a variable-temperature cryostat at 100 and 300~K.
The diffuse scattering data was converted reciprocal space using the CrysAlisPro software.

To compare to these experimental measurements, we have calculated the theoretical diffuse scattering from our DFT calculations by calculating the difference in the scattering between the pristine and defective unit cells, following the approach in Ref.~\cite{hutchings1984investigation}.
This approach yields diffuse scattering that is valid away from the Bragg peaks (which are not present in this calculation), whilst avoiding the need to use a supercell which would be too large to relax using DFT.
In the case of the lab-based x-ray measurements, a $q$-independent background has been added to the calculations to provide a better comparison to the data.

\subsection{Powder diffraction}
Powder x-ray diffraction measurements of \pzo\ were performed using a Bruker D8 DISCOVER XRD system at Royal Holloway, University of London, UK, and subsequently refined using the TOPAS software.
The measurements were performed in Bragg-Brentano focusing geometry, with a fixed sample illumination.
All measurements were performed at room temperature.

The time-of-flight neutron diffraction was measured using the Polaris high-flux, medium resolution diffractometer at the ISIS Neutron and Muon Source, UK~\cite{smith2019upgraded}.
The $\approx10$~g powder sample was loaded into a thin-walled cylindrical vanadium can, and data were collected at room temperature for a duration of 8 hours.
Data reduction and generation of files suitable for profile refinement used the Mantid open-source software~\cite{arnold2014mantid}.
The data from all five detector banks were fitted simultaneously with the TOPAS Rietveld refinement program.

\subsection{DC and AC magnetic susceptibility}
DC susceptibility measurements of both single crystal and powder samples of \pzo\ were performed using a Quantum Design MPMS3 system at the University of Oxford, and using a similar system at the ISIS Neutron and Muon Source Materials Characterisation Laboratory, UK.
Similarly, measurements of the AC susceptibility were performed using a Quantum Design PPMS DynaCool with the ACMS II option and a dilution fridge inset at the ISIS Neutron and Muon Source Materials Characterisation Laboratory, UK.
The single crystal sample was mounted on a quartz rod whose response had already been measured to allow correct background removal.
The magnetic field was applied such that it was parallel to the [111] crystallographic direction, and measurements were always performed after cooling from high temperature in zero-applied field.

\subsection{Inelastic neutron scattering}
Inelastic neutron scattering measurements of powder \pzo\ were performed using the SEQUOIA instrument~\cite{granroth2010sequoia} at the Spallation Neutron Source, Oak Ridge National Laboratory, USA.
Data were collected with 25, 60, and 120~meV incident energies.
The phonon spectrum from the non-magnetic analogue La$_2$Zr$_2$O$_7$ was also measured, which was subsequently subtracted, along with effects from the Al sample can, from the \pzo\ data.
Data were normalised to the 25 meV data by assuming that the total elastic scattering was the same for each incident energy.
The remaining features were checked to see if they were magnetic in origin by looking at the $\left|\vect{Q}\right|$ dependence, and comparing it to the expected dependence, i.e., that of the Pr$^{3+}$ magnetic form factor.
The data were analysed using the Mantid package~\cite{arnold2014mantid}.

\subsection{Point charge model}
A point charge model of the crystal electric field environment around the Pr ions was fed with the positions obtained from the DFT calculations.
For the pristine structure the high symmetry means that all 16 Pr atoms in the conventional unit cell are equivalent and have real eigenkets, whereas in the defect structure all 16 Pr atoms are symmetrically inequivalent and possess complex eigenkets.
The calculation of the point charge model, and subsequent physical properties (inelastic neutron scattering spectra, magnetometry, heat capacity), was carried out using the PyCrystalField python library~\cite{scheie2021pycrystalfield}.

\subsection{Magnetic diffuse scattering simulations}
We model \pzo\ through the classical nearest-neighbour spin ice Hamiltonian $H = J\sum_{\langle ij\rangle} S_i S_j,$ where $S_i=\pm1$ are Ising variables on each pyrochlore site.
Estimates for the coupling strength $J$ in the literature~\cite{kimura2013quantum,wen2017disordered,tokiwa2018discovery} range between 0.8~K and 1.7~K in this convention; for concreteness, we take $J=1$~K.
Monte Carlo simulations were carried out using the single-spin-flip Metropolis algorithm and simulated annealing, using the codebase developed for Ref.~\cite{pearce2022magnetic}, available at \url{https://github.com/attila-i-szabo/Ho2Ir2O7/tree/v1.0}, which natively generates reciprocal-space correlation functions.
We implemented diluted spin ice by permanently removing each spin independently with probability 30\% at random before starting the Monte Carlo steps.
The neutron scattering patterns in the main text were obtained by averaging over 32 different configurations of removed sites and 8 simulation histories for each configuration in the diluted case, and over 256 simulation histories in the pristine case.

\begin{figure*}
	\centering
	\includegraphics[width=\linewidth]{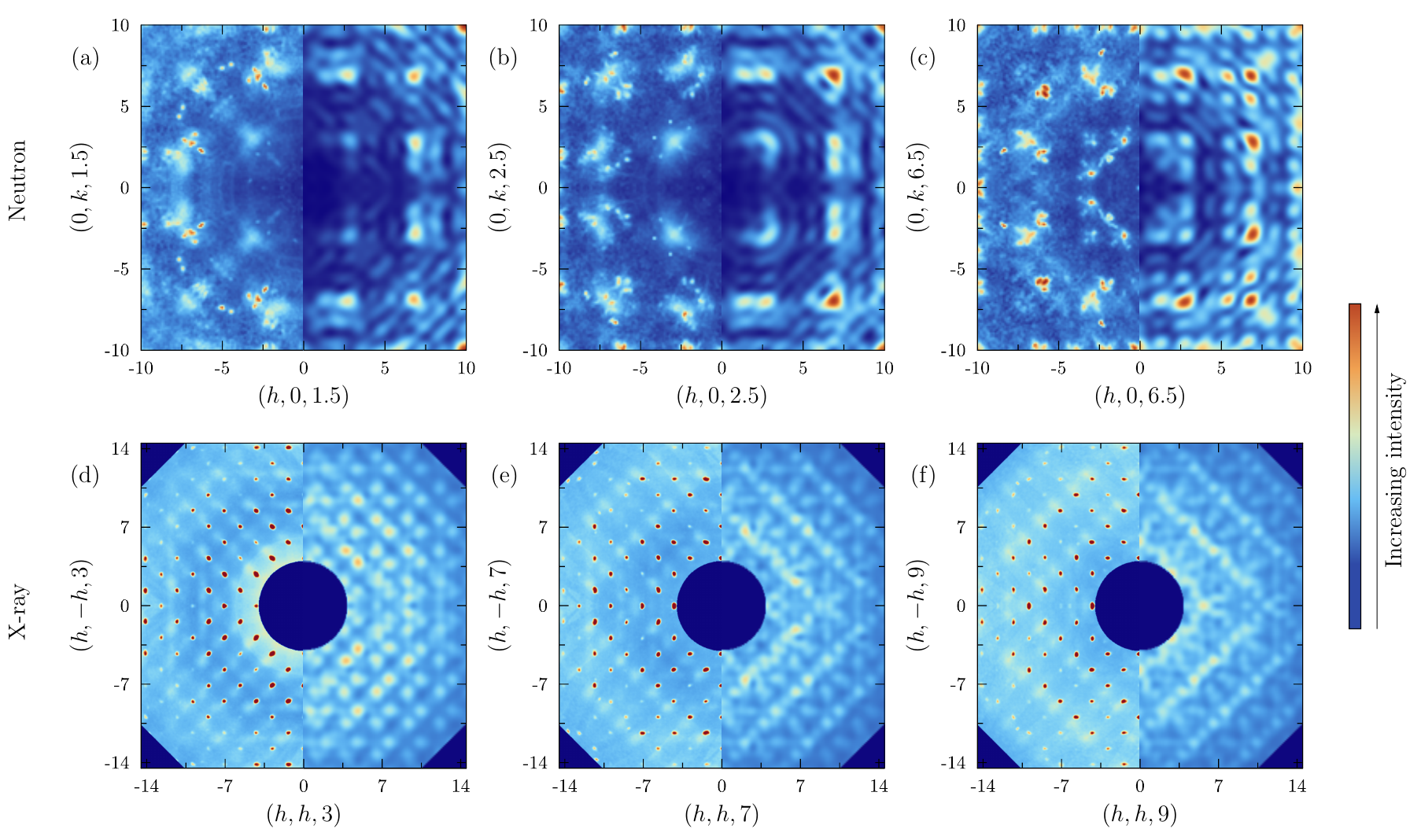}
	\caption{\textbf{Additional diffuse scattering measurements and calculations}. On the top (a--c) we show two extra planes of diffuse neutron scattering from \pzo, and on the bottom (d--f), similar additional planes from the diffuse x-ray scattering are presented. In all cases, they are compared to calculations, as described in the main text, with the calculations shown in the right-hand half of each panel.}
	\label{fig:extraDiffuse}
\end{figure*}

\section{Additional diffuse scattering}
In the main text we show the good match between our experimentally measured diffuse scattering, and the predictions based on our DFT calculations.
We show a similarly good match in some additional planes in Fig.~\ref{fig:extraDiffuse}, shown for both neutron and x-ray measurements.
For completeness, we also repeat the planes from the main text.
The good match demonstrates that the defect model discussed in the main text produces diffuse scattering that well describes the experimental measurements over the entire measured $Q$ range.
As in the main text, in the case of the x-ray diffuse scattering calculations, a constant background has been added to the calculations to provide a better comparison to the data.

\begin{figure}
	\centering
	\includegraphics[width=\linewidth]{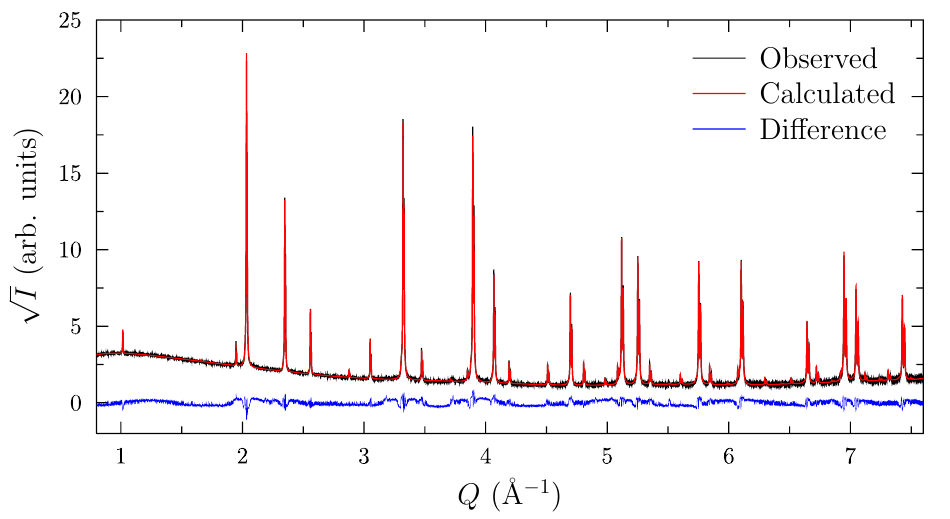}
	\caption{\textbf{Powder x-ray diffraction measurements of \pzo}. The powder-xray diffraction profile is shown, along with the best Reitveld refinement. The difference between the model and data is shown in blue.}
	\label{fig:pxrd}
\end{figure}

\section{Rietveld refinements}
\subsection{Powder x-ray diffraction}
In the main text we introduce powder x-ray diffraction measurements, with the corresponding fit presented in Fig.~\ref{fig:pxrd}.
In these measurements, there is insufficient sensitivity to be able to accurately extract the signature of the defect structures, specifically occupation of Pr$^{3+}$ atoms on the Zr site, and O vacancies/interstitials.
As such, we have fit the data to the pristine structure in space group 227, obtaining a lattice parameter of 10.700496(7)~\AA, which is within the range reported in Refs.~\cite{kimura2013quantum,hatnean2014structural,koohpayeh2014synthesis}.
The refined fractional $x$ coordinate of the O ion sitting on the 48f site is 0.08472(17).
For this refinement we have a weighted R-factor $R_\text{wp}~=~4.82$\%, with this fit to the data shown in Fig.~\ref{fig:pxrd}.

\begin{figure}[p]
	\centering
	\includegraphics[width=0.85\linewidth]{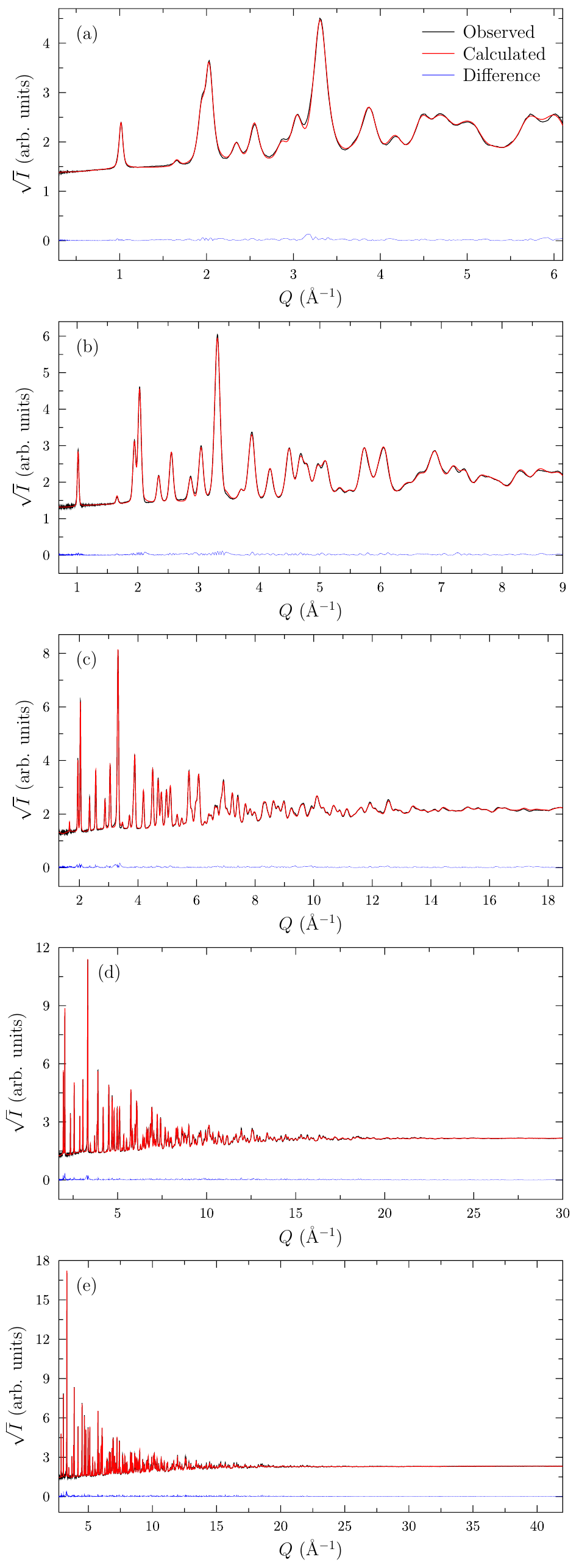}
	\caption{\textbf{Powder neutron diffraction measurements of \pzo}. The powder neutron diffraction patterns are shown, with five different $Q$ ranges corresponding to the different detector banks on the Polaris instrument. The best simultaneous Reitveld refinement, as detailed in Tabs.~\ref{tab:pndi} and \ref{tab:pndii}, and the difference between the model and data, are shown on each.}
	\label{fig:pnd}
\end{figure}

\begin{table}
	\begin{tabular}{c|c|ccc}
		\textbf{Atom type} & \textbf{Wyckoff position} & \textit{\textbf{x}} & \textit{\textbf{y}} & \textit{\textbf{z}} \\ \hline
		Pr$_{16\text{d}}$ & 16d & 0.5 & 0.5 & 0.5 \\
		Pr$_{16\text{c}}$ & 16c & 0 & 0 & 0 \\
		Zr & 16c & 0 & 0 & 0 \\
		O$_{8\text{b}}$ & 8b & 0.375 & 0.375 & 0.375 \\
		O$_{48\text{f}}$ & 48f & $x$ & 0.875 & 0.875 \\
		O$_{8\text{a}}$ & 8a & 0.875 & 0.875 & 0.875
	\end{tabular}
	\caption{\textbf{Crystallographic sites used for Rietveld refinements to powder neutron measurements of \pzo}. The position of each atom (space group 227, origin choice 2) is given in fractional coordinates, as well as the corresponding Wyckoff site. For each of the different fits, the fractional occupation $f$ and isotropic thermal parameter $B$ are given in Table~\ref{tab:pndii}.}
	\label{tab:pndi}
\end{table}

\begin{table*}
	\begin{tabular}{ccc||ccc||ccc}
		\multicolumn{3}{c||}{\textbf{Pristine}} & \multicolumn{3}{c||}{\textbf{DFT}} & \multicolumn{3}{c}{\textbf{Free}} \\ \hline\hline
		\textbf{\textit{\textbf{R$_\text{wp}$~$(\text{\%})$}}} & \textbf{\textit{\textbf{a~$(\text{\AA})$}}} & \textbf{\textit{\textbf{x}}} & \textbf{\textit{\textbf{R$_\text{wp}$~$(\text{\%})$}}} & \textbf{\textit{\textbf{a~$(\text{\AA})$}}} & \textbf{\textit{\textbf{x}}} & \textbf{\textit{\textbf{R$_\text{wp}$~$(\text{\%})$}}} & \textbf{\textit{\textbf{a~$(\text{\AA})$}}} & \textbf{\textit{\textbf{x}}} \\ \hline
		3.40 & 10.700(2) & 0.08458(4) & 3.28 & 10.702(4) & 0.08450(4) & 3.25 & 10.700(4) & 0.08453(4) \\ \hline
		\textbf{Atom} & \textbf{\textit{\textbf{f}}} & \textbf{\textit{\textbf{B~(\AA$^2$)}}} & \textbf{Atom} & \textbf{\textit{\textbf{f}}} & \textbf{\textit{\textbf{B~(\AA$^2$)}}} & \textbf{Atom} & \textbf{\textit{\textbf{f}}} & \textbf{\textit{\textbf{B~(\AA$^2$)}}} \\ \hline
		Pr$_{16\text{d}}$ & 1 & 0.413(11) & Pr$_{16\text{d}}$ & 1 & 0.471(11) & Pr$_{16\text{d}}$ & 1 & 0.458(8) \\
		Pr$_{16\text{c}}$ & - & - & Pr$_{16\text{c}}$ & 0.052(2) & 0.471(11) & Pr$_{16\text{c}}$ & 0.055(6) & 0.458(8) \\
		Zr & 1 & 0.302(7) & Zr & 0.948(2) & 0.270(7) & Zr & 0.945(6) & 0.264(7) \\
		O$_{8\text{b}}$ & 1 & 0.301(11) & O$_{8\text{b}}$ & 1 & 0.370(11) & O$_{8\text{b}}$ & 1 & 0.358(11) \\
		O$_{48\text{f}}$ & 1 & 0.582(7) & O$_{48\text{f}}$ & 0.9827(5) & 0.573(6) & O$_{48\text{f}}$ & 0.992(4) & 0.587(9) \\
		O$_{8\text{a}}$ & - & - & O$_{8\text{a}}$ & 0.052(2) & 0.370(11) & O$_{8\text{a}}$ & 0.046(4) & 0.358(11)
	\end{tabular}
	\caption{\textbf{Refined parameters from Rietveld refinements to powder neutron measurements of \pzo}. For all three different models (described in the text), we provide the goodness-of-fit $R_\text{wp}$, lattice parameter $a$, position of O on the 48f Wyckoff site $x$ (see Table~\ref{tab:pndi}), site occupancies $f$, and isotropic thermal parameters $B$.}
	\label{tab:pndii}
\end{table*}

\subsection{Powder neutron diffraction}
In contrast to the powder x-ray measurements above, the powder neutron measurements are able to detect changes in the patterns that correspond to the defect structure.
On the Polaris instrument, the data is collected in five banks of detectors, each of which is sensitive to a different $Q$ range.
One such bank is shown in the main text; we show the entire set here in Fig.~\ref{fig:pnd}.
All fits described below simultaneously fit the data in all banks, and we have once again used space group 227.
For these data, we have trialled three different models: a pristine structure with no defects, the model derived using DFT, and a relatively free fit of occupancies based on this model.
The details of the fit parameters for these data are shown in Tables~\ref{tab:pndi} and \ref{tab:pndii}.
In the case of the DFT model, the occupation numbers are constrained such that $f_{\text{Pr}_{16\text{c}}}$ is fitted, $f_\text{Zr} = 1 - f_{\text{Pr}_{16\text{c}}}$,  $f_{\text{O}_{48\text{f}}} = 1 – f_{\text{Pr}_{16\text{c}}}/3$, and $f_{\text{O}_{8\text{a}}} = f_{\text{Pr}_{16\text{c}}}$.
In the case of the free model, $f_{\text{Pr}_{16\text{c}}}$, $f_{\text{O}_{48\text{f}}}$ and $f_{\text{O}_{8\text{a}}}$ are fitted and $f_\text{Zr} = 1 - f_{\text{Pr}_{16\text{c}}}$.
In both cases, the thermal parameters for the stuffed and interstitial sites are constrained such that $B_{\text{Pr}_{16\text{c}}} = B_{\text{Pr}_{16\text{d}}}$ and $B_{\text{O}_{8\text{a}}} = B_{\text{O}_{8\text{b}}}$.
We note that the defect models give a better fit to the data than the pristine model.
Furthermore, when the occupancies are allowed to vary, the fitted values are consistent with the DFT model.
The greater $Q$ range available with neutrons allows the thermal parameters to be determined more accurately than with x-rays, and the fact that, when compared to the x-ray refinements, these fits have lower thermal parameters on the O sites adds confidence to the validity of the defect model.

Alternative models were trialled, however this was the best match to the data.
Specifically, there was no evidence of any Zr on the Pr (16d) site.
The DFT results presented in the main text suggest the O interstitial is actually displaced away from the 8a site towards the 32e site, however putting the O interstitial on the 32e site does not improve the quality of the fit.

\begin{table}
	\begin{tabular}{c|cccc}
		\textbf{Atom} & \textit{\textbf{$U_{11}$}} & \textit{\textbf{$U_{22}=U_{33}$}} &  \textit{\textbf{$U_{12}=U_{13}$}} &  \textit{\textbf{$U_{23}$}} \\ \hline
		Pr$_{16\text{d}}$ & 0.0150(11) & 0.0150 & -0.0043(2) & -0.0043 \\
		Zr$_{16\text{c}}$ & 0.0107(13) & 0.0107 & 0.0008(2) & 0.0008 \\
		O$_{8\text{b}}$ & 0.011(3) & 0.011 & 0 & 0 \\
		O$_{48\text{f}}$ & 0.029(4) & 0.014(3) & 0 & 0.005(3) \\
	\end{tabular}
	\caption{\textbf{Anisotropic thermal displacement parameters $U$ from refinement of the single crystal x-ray diffraction from \pzo}. The $U$ values are expressed in units of \AA$^2$, and the atom type is defined in Table~\ref{tab:pndi}. The fit assumes full occupation of sites for space group 227, and we obtain a lattice parameter $a$ = 10.7418(4)~\AA, a fractional coordinate $x$ = 0.0868(8) (defined in Table~\ref{tab:pndi}). For Pr$_{16d}$ and Zr$_{16c}$ we set $U_{11} = U_{22} = U_{33}$ and $U_{12} = U_{13} = U_{23}$, for O$_{8b}$ we set $U_{11} = U_{22} = U_{33}$ and $U_{12} = U_{13} = U_{23} = 0$, and for O$_{48f}$ we set $U_{22} = U_{33}$ and $U_{12} = U_{13} = 0$, by symmetry.}
	\label{tab:sxxd}
\end{table}

\begin{figure}
	\centering
	\includegraphics[width=0.7\linewidth]{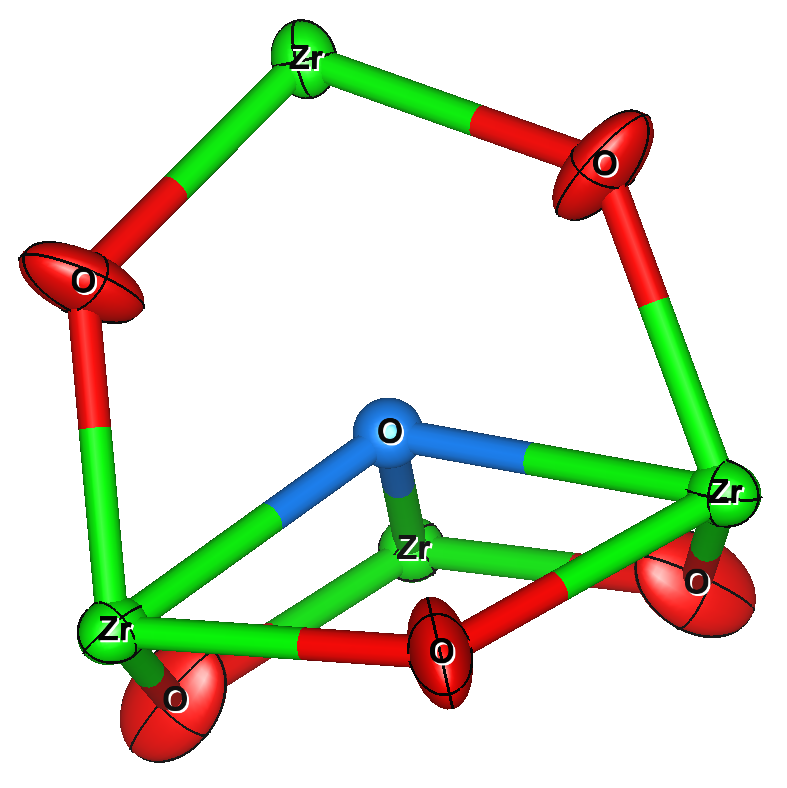}
	\caption{\textbf{Anisotropic thermal ellipsoids extracted from x-ray diffraction our single crystal sample of \pzo}. Including the anisotropic thermal parameters significantly improves the quality of the fit, with the extension of the O 48f ions (red) towards the O 8a site (blue) matching our DFT predictions.}
	\label{fig:aniostropicThermalParameters}
\end{figure}

\subsection{Single crystal x-ray diffraction}
The refinement of the single-crystal x-ray diffraction data had no sensitivity to oxygen vacancies or interstitials.
However, there is a large reduction in the $R$-factor when going from isotropic ($R_\text{wp}=7.8\%$) to anisotropic thermal displacement parameters ($R_\text{wp}=2.7\%$).
The fitted anisotropic thermal displacement parameters are shown in Table~\ref{tab:sxxd}.
A graphical representation of the anisotropic thermal ellipsoids is shown in Fig.~\ref{fig:aniostropicThermalParameters}.
The elongation of the O 48f ellipsoids along the direction to the nearest O 8a interstitial site agrees with the predictions of our DFT geometry optimisation.

\begin{table*}
	\begin{tabular}{c||c|c|c|c|c|c|c}
		\textbf{Parameter} & \textbf{Pr excess} & \textbf{Frenkel} & \textbf{O$_\text{8a}$ interstitial} & \textbf{Zr excess} & \textbf{O$_\text{8b}$ vacancy} & \textbf{O$_\text{48f}$ vacancy} & \textbf{Pristine} \\
		\hline\hline
		$R_\text{wp}$~(\%) & 3.25 & 3.25 & 3.27 & 3.27 & 3.37 & 3.38 & 3.40 \\
		$f(\text{Pr}_\text{16d})$ & 1 & 1 & 1 & 1.000(4) & 1 & 1 & 1 \\
		$f(\text{Pr}_\text{16c})$ & 0.055(6) & 0 & 0 & 0 & 0 & 0 & 0 \\
		$f(\text{Zr}_\text{16c})$ & 0.945(6) & 1 & 1 & 1 & 1 & 1 & 1 \\
		$f(\text{Zr}_\text{16d})$ & 0 & 0 & 0 & 0.000(4) & 0 & 0 & 0 \\
		$f(\text{O}_\text{8b})$ & 1 & 1 & 1 & 1 & 1.058(8) & 1 & 1 \\
		$f(\text{O}_\text{48f})$ & 0.992(4) & 1.005(4) & 1 & 1 & 1 & 0.999(4) & 1 \\
		$f(\text{O}_\text{8a})$ & 0.046(4) & 0.051(8) & 0.049(8) & 0.050(7) & 0 & 0 & 0 \\
	\end{tabular}
	\caption{\textbf{Refined parameters from Rietveld refinements of different defect models to powder neutron measurements of \pzo}. For all different models, we provide the goodness-of-fit $R_\text{wp}$ and site occupancies $f$.}
	\label{tab:altModels}
\end{table*}

\section{Alternative defect models}
Alternative defect models were trialled and excluded.
We start by comparing the results of Rietveld refinements of different defect models, as shown in Tab.~\ref{tab:altModels}.
In particular, we compare defect models that have been suggested for other similar pyrochlore systems, such as Frenkel disorder~\cite{sibille2017coulomb}, where there are vacancies on the 48f O site, and charge compensating O 8a interstital sites are occupied (i.e. similar to our model, but without the Pr ions on Zr sites).
As can be seen, the goodness-of-fit parameter $R_\text{wp}$ clearly indicates that some models are better than others.
The O vacancy and pristine models were rejected on the basis of their significantly higher $R_\text{wp}$.
In fact, in the case of the O vacancy models, the fitted occupations do not suggest oxygen deficiency.
Similarly, the Zr excess model does not suggest any excess Zr ions on Pr sites.
The Frenkel model gives occupation of the O interstitial site, but no charge compensating O vacancies.
The Pr excess model in the main paper gives the lowest $R_\text{wp}$ and site occupations that are consistent with charge neutrality.

\begin{figure*}
	\centering
	\includegraphics[width=\linewidth]{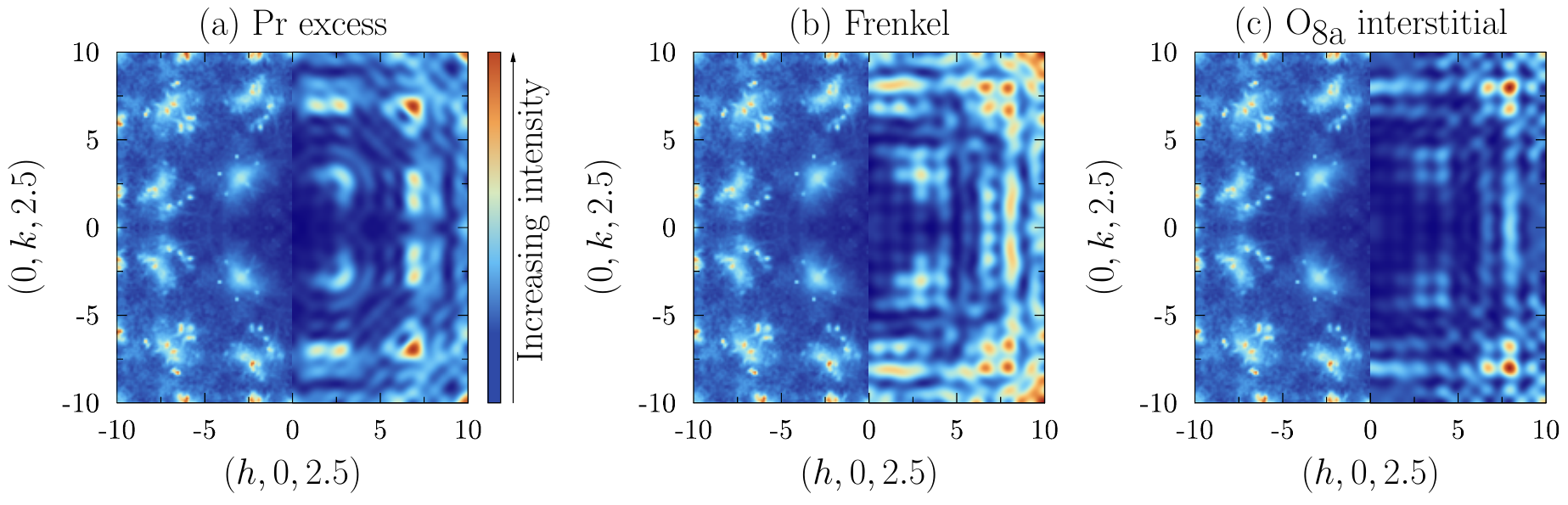}
	\caption{\textbf{Diffuse scattering calculations for alternative defect models}. In all three panels, on the left we show the experimentally measured diffuse neutron scattering. On the right we present three different models, all of which use the DFT-calculated atomic positions: (a) Pr excess, i.e. the model in the main text, (b) the Frenkel model, and (c) an interstitial O on the 8a site.}
	\label{fig:altModels}
\end{figure*}

To further compare the models, one must also perform the DFT calculations and calculate the crystallographic diffuse scattering, as done for the defect model presented in the main text.
As these calculations are very expensive, we only perform them for the two most promising additional possible defect structures (i.e. those with the lowest $R_\text{wp}$), specifically Frenkel defects and interstitial O on the 8a site.
The diffuse scattering in the $(h, k, 2.5)$ plane (the same as shown in the main text) is shown in Fig.~\ref{fig:altModels}.
The best agreement is clearly the Pr excess model presented in the main text.

\begin{figure}
	\centering
	\includegraphics[width=\linewidth]{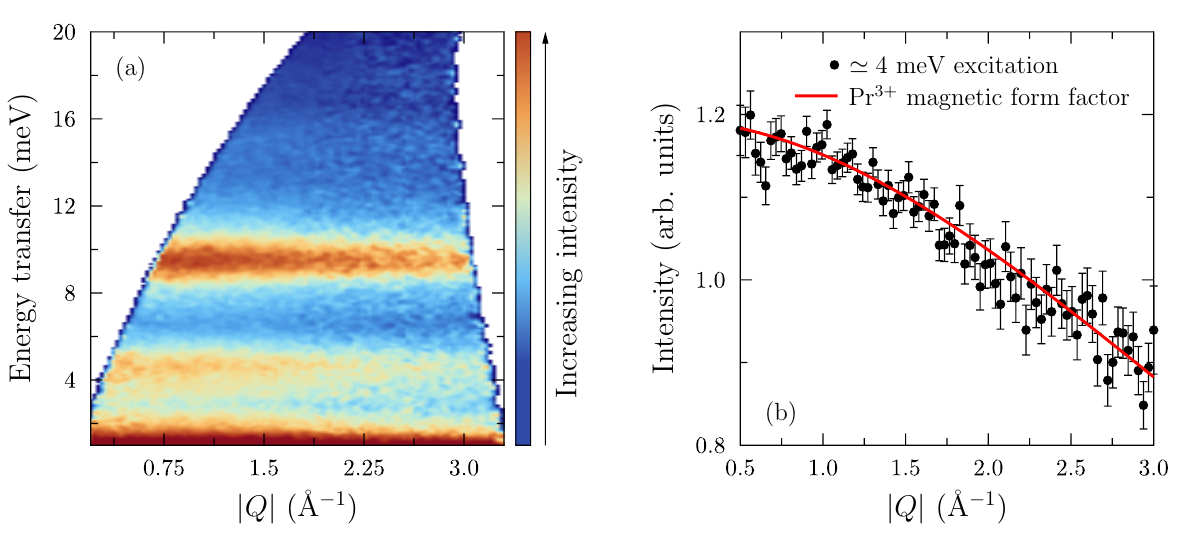}
	\caption{\textbf{Raw inelastic neutron scattering data from \pzo}. (a) $S\left(Q,\omega\right)$ measured with an incident energy of 25~meV at a temperature of 5~K after subtraction of the scattering from the phonon blank La$_2$Zr$_2$O$_7$. The $Q$-dependence of the scattering at around 4~meV is shown in (b).}
	\label{fig:formFactor}
\end{figure}

\section{Inelastic neutron scattering}
The $Q$-dependence of all peaks in the inelastic neutron scattering data presented in the main text was checked to confirm whether the feature was magnetic in origin (or arises instead from a phonon-like feature).
In Fig.~\ref{fig:formFactor} we show such a check for the $\simeq4$~meV excitation that we identify as the lowest lying crystal field excitation in the defective \pzo.
The intensity clearly drops following the magnetic form factor of Pr$^{3+}$, confirming its magnetic origin.

\begin{figure}
	\centering
	\includegraphics[width=\linewidth]{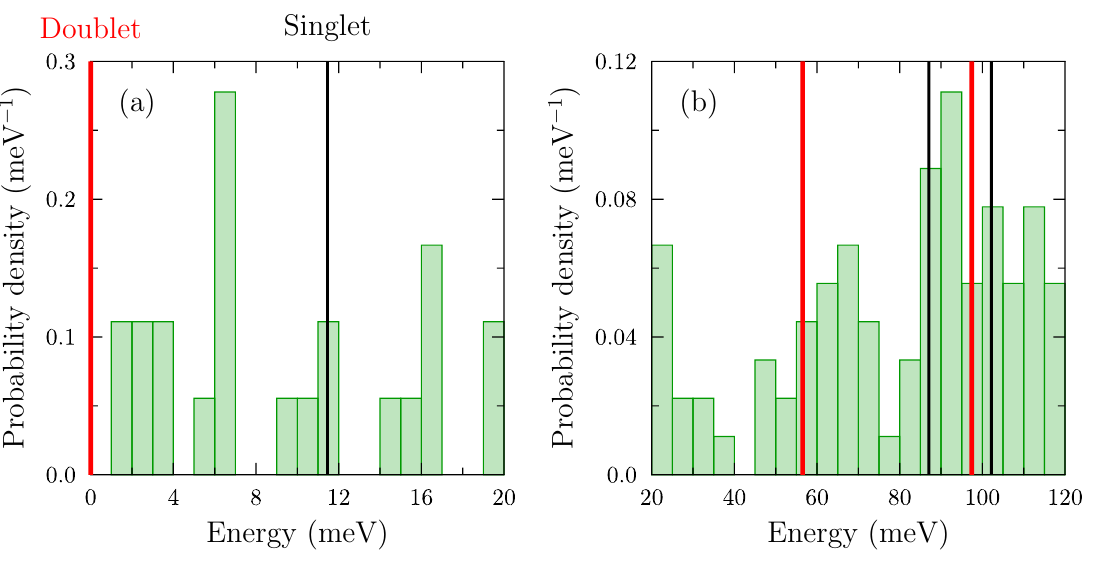}
	\caption{\textbf{The probability of finding a crystal field excitation at a certain energy in the defective unit cell}. The crystal electric field of \pzo\ was calculated by a point-charge model, informed by the atomic positions from our density functional theory calculations. The defect structure gives a continuum of singlets (green bars). For comparison the well defined doublets (red solid lines) and singlets (black solid lines) of pristine \pzo\ are marked.}
	\label{fig:energyLevels}
\end{figure}

\section{Point charge model}
In addition to the predictions of the inelastic neutron scattering spectra shown in the main text, an alternative way to inspect the results of the point charge model of the intrinsically defective \pzo\ structure, which takes the atomic positions from our density functional theory calculations, is by looking at the energy levels.
As the defect structure leads to 18 inequivalent Pr sites (16 on the standard Pr site, plus two Zr sites that have Pr occupying them), rather than a well defined set of energy levels, we instead have a relatively continuous set.
These are visualed in Fig.~\ref{fig:energyLevels}, where we observe that there doublet and singlet energy levels of pristine \pzo\ have given way to a more complex picture.
It is these complex landscape of energy levels that gives rise to the inelastic neutron scattering results shown in the main text.

\begin{figure}
	\centering
	\includegraphics[width=0.7\linewidth]{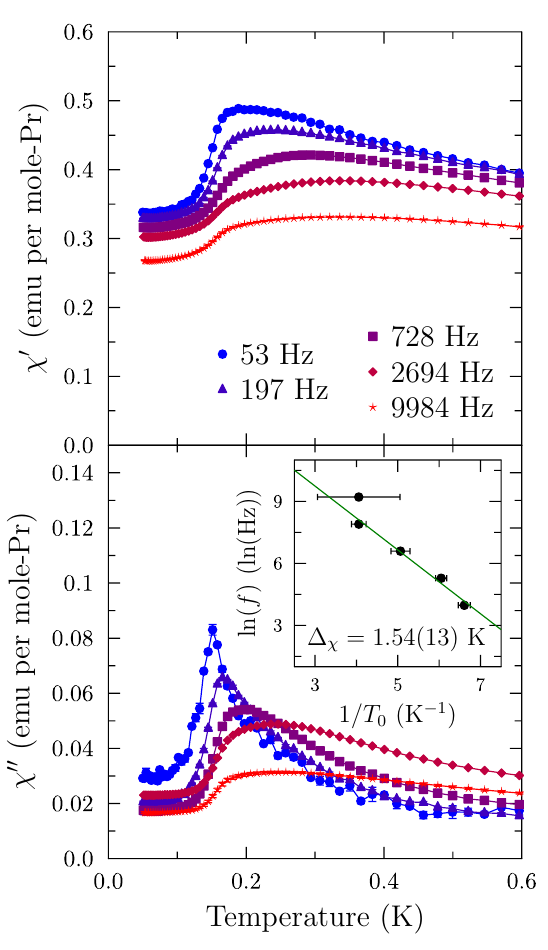}
	\caption{\textbf{The AC susceptibility response of our single crystal sample of \pzo}. The real (top) and imaginary (bottom) parts are shown. The frequency dependence of the peak in the imaginary part is shown as an inset, and demonstrates Arrhenius behaviour. The magnetic field was applied such that it was parallel to the [111] crystallographic direction.}
	\label{fig:acsus}
\end{figure}

\section{AC and DC susceptibility}
Measurements of the (inverse) magnetic susceptibility provide further evidence that our samples behave similarly to previous reports~\cite{matsuhira2009spin,kimura2013quantum,hatnean2014structural,koohpayeh2014synthesis,alam2016crystal,bonville2016magnetic,petit2016antiferroquadrupolar,wen2017disordered,tokiwa2018discovery,tang2020magnetization,tang2023spin}.
As discussed in the main text, the Curie-Weiss temperature of our sample is estimated to be around 1.5~K by fitting the data below 10~K.
We note that accurately estimating the Curie-Weiss temperature in \pzo\ is very challenging, especially when considering the complicated crystal electric field effects that are present, and the significant curvature of the inverse magnetic susceptibility.
As such, we suspect that at least some of the variance in this parameter that has been reported in the literature is due to these challenges.
In Ref.~\cite{tang2023spin} a correspondence between the Curie-Weiss temperature and crystallographic disorder is suggested; specifically, samples with a Curie-Weiss temperature closer to zero appear to have less crystallographic disorder.
Within this framework, our sample is therefore typical of those in the majority of the literature where quantum spin ice behaviour has been reported.
We calculated the susceptibility from the model combining pristine and defect structures in the proportions found in our INS measurements, finding that our defect structure has minimal impact upon the contribution to the bulk susceptibility that arises purely from single-ion physics.
We therefore conclude that magnetic susceptibility does not allow one to directly draw any strong conclusions on the defect structure or its effect on the magnetism without reference to other measurements.
Other magnetic measurements, such as the $M(H)$ curves collected at mK temperatures in Ref.~\cite{tang2023spin}, may be more informative.

As discussed in the main text, our \pzo\ sample has a similar magnetic response to other reports, most notably that of Ref.~\cite{kimura2013quantum}, where AC susceptibility measurements are reported.
We show similar measurements on our sample in Fig.~\ref{fig:acsus}, where we demonstrate the same behaviour.
Most notably, the frequency dependent peak, which was previously explained as the thermal activation of spin flips in the spin ice material, follows the same Arrhenius law dependency with the sample energy gap.

\begin{figure}
	\centering
	\includegraphics[width=0.7\linewidth]{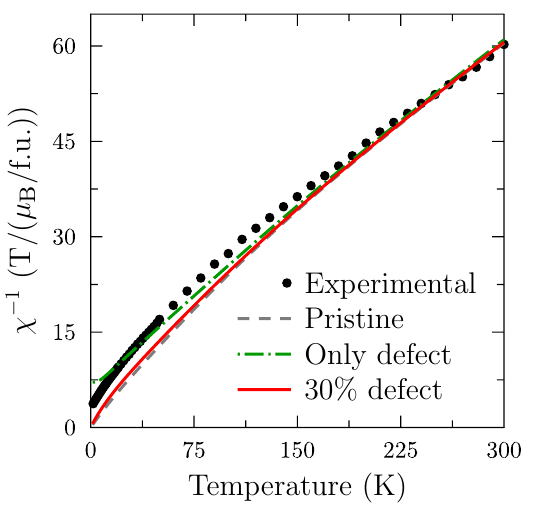}
	\caption{\textbf{The DC susceptibility response of our powder sample of \pzo}. The combination of the pristine cell with the defect structure can be used to predict the magnetisation using the point charge model.}
	\label{fig:dcsus}
\end{figure}

We can also explore the DC susceptibility of \pzo.
Our measurements, see Fig.~\ref{fig:dcsus} of the (inverse) magnetic susceptibility show an abnormal curvature, as also reported by many others~\cite{matsuhira2009spin,kimura2013quantum,hatnean2014structural,koohpayeh2014synthesis,alam2016crystal,bonville2016magnetic,tang2020magnetization}.
We can model these data using the point charge model discussed in the main text by calculating the susceptibility from the model combining pristine and defective structures in the proportions found in our INS measurements.
Whilst the effects are small, we find that this abnormal curvature is enhanced by the inclusion of the defect structure, as is the non-zero intercept.
Our match to the experimentally observed inverse magnetic susceptibility is improved by the inclusion of the defect structure for all single crystal orientations and in powder, however given the uncertainty in the measurements and small theoretical change, one must be careful not to over-interpret this result.

\begin{figure}
	\centering
	\includegraphics[width=0.7\linewidth]{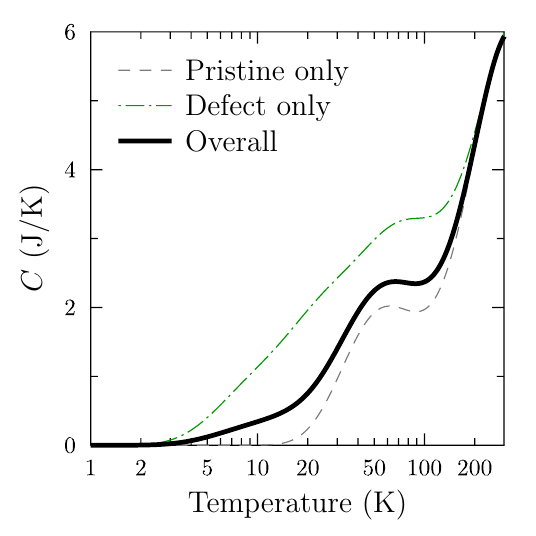}
	\caption{\textbf{The contribution to the heat capacity of \pzo\ arising from crystal field effects}. This contribution is simulated from our DFT-inspired point charge model. We show the difference between pristine and defective \pzo. The overall curve, i.e. the prediction for \pzo\ when the defects are accounted for, is simply the sum of the pristine only and defect only curves in the proportion found in the main text.}
	\label{fig:heatcapacity}
\end{figure}

\section{Heat capacity}
In Ref.~\cite{kimura2013quantum} the heat capacity of \pzo\ is analysed to learn more about the magnetism of the system.
For this to be done, the contribution to the heat capacity that is believed to arise from the single-ion magnetism is subtracted from the measured data.
However, as we have now demonstrated, this is based on an incomplete model of the crystal electric field effects in \pzo\ which ignores the intrinsic defect structure.
We show the simulated crystal field contribution to the heat capacity for pristine \pzo, and when the defects are included, in Fig.~\ref{fig:heatcapacity}.
As can clearly be seen, there is a non-negligible contribution to the heat capacity at an order-of-magnitude lower temperature when the defect structure is considered compared to when the pristine structure is considered alone.
Specifically, without considering the defect structure of \pzo, one finds that there is no significant contribution to the heat capacity from the crystal electric field effects below $T\simeq15$~K; inclusion of the realised defect structure results in significant contributions to the heat capacity at $T\simeq2.5$~K and higher.
These temperatures coincide with the temperatures where Ref.~\cite{kimura2013quantum} identified a broad peak in the heat capacity that they link to the proliferation of thermally activated magnetic monopole pairs; this conclusion must be revisited in light of our findings.
As with the interpretation of the magnetic diffuse scattering, these simulations do not exclude the possibility of the quantum spin ice state, however these effects arising from intrinsic defects need to be accounted for in order to extract quantitative information from the system.

\section{Additional notes on the Monte Carlo simulations}
According to our work, many of the Pr ions in \pzo\ have local environments whereby the low-lying single-ion crystal electric field energy levels are non-magnetic singlets, with a gap between the ground state and the first excited state considerably larger than the exchange, dipolar, or other interactions. 
While interactions always introduce some mixing between the single-ion CEF states, and therefore the capability to develop a non-vanishing magnetic moment, in this setting the mixing is perturbatively small and so is the moment. 
Treating such ions in the conventional way of a doublet weakly split by a transverse field (sometimes called a `quasi-doublet') is not a good approximation. In fact, as we show below, they can be sensibly regarded as non-magnetic ions to leading order. 

Consider a two-level system with $H_x$, a transverse-field that arises from the defect structure, creating the gap between the two energy levels, and $H_z \ll H_x$, a longitudinal-field that weakly mixes the two states, arising from the interactions with other spins in the system.
The Hamiltonian operator is thus
\begin{equation}
	\mathcal{\hat{H}} = \frac{1}{2}\begin{pmatrix}
		H_z & H_x \\
		H_x & -H_z
	\end{pmatrix} = \frac{H_x}{2}\sigma_x + \frac{H_z}{2}\sigma_z ,
\end{equation}
where $\sigma_i$ are the standard Pauli matrices ($\hbar = 1$). 
The eigenvalues of this system are $E_{\pm} = \pm\frac{1}{2}\sqrt{H_x^2+H_z^2}$, where $E_-$ is the energy of the ground state, and $E_+$ is the energy of the excited state.
Their difference in energy is $\Delta=E_+-E_-$, and one can recast the unknown $H_x$ in terms of the experimentally measured gap size $\Delta$: $H_x=\sqrt{\Delta^2-H_z^2}$.

We utilise the Hellmann-Feynmann theorem to state that
\begin{equation}\label{eqn:hellmann-feynmann}
	\left<\psi_-\left|\frac{\partial\mathcal{\hat{H}}}{\partial H_z}\right|\psi_-\right> = \frac{\partial E_-}{\partial H_z},
\end{equation}
where $\psi_-$ is the ground state of the system.
As
\begin{equation}
	\frac{\partial\mathcal{\hat{H}}}{\partial H_z} = \frac{1}{2}\sigma_z = S_z ,
\end{equation}
we can evaluate Eq.~\ref{eqn:hellmann-feynmann} to show that
\begin{equation}
	\left<S_z\right> = -\frac{H_z}{2\sqrt{H_x^2+H_z^2}} = -\frac{H_z}{2\Delta}.
\end{equation}
This scales as expected: when $H_z=0$ (i.e., in the absence of any interactions), $\left<S_z\right>=0$, and when $H_x=0$ (i.e., in the absence of disorder), $\left<S_z\right> = -1/2$.

We can write an equation for the expected magnetic moment of a defective Pr in \pzo, given the value of the gap $\Delta$, as
\begin{equation}\label{eqn:gappedMoment}
	m_\text{Pr}^\text{defect} = \frac{m_\text{Pr}H_z}{\Delta} ,
\end{equation}
where $m_\text{Pr}$ is the moment of the pristine Pr moment with a doublet ground state.
We can finally calculate the approximate low-temperature magnetic moments of the defective Pr ions using the gap sizes from our point charge model of the DFT-optimised defective unit cell, which values between $\simeq1.2$~meV and $\simeq20$~meV, with an average gap size of around $7.5$~meV.
The value of $H_z$ has been reported to be anywhere between 0.5~K~\cite{martin2017disorder} and 1.7~K~\cite{tokiwa2018discovery} (with the smaller value coming from systems where disorder is taken into account).
Using Eq.~\eqref{eqn:gappedMoment}, we thus find that the maximum moment predicted by this two-level system calculation for any of the 18 Pr ions in the defective unit cell (corresponding the small gap) is somewhere between 4\% and 13\% of the full Pr moment.
The average moment in the cell is much lower, between 1\% and 4\%, with the minimum moment (corresponding to the largest gap) being between 0.2\% and 0.7\% of the full Pr moment.
Inclusion of higher energy levels allows further mixing between levels and a slight increase in the magnetic moment, but does not change the result in a significant way. Hence, we conclude that the assumption made in the main text, specifically that all the Pr moments in the defective unit cell can be approximated as having zero magnetic moment in the Monte Carlo calculations, is well justified.

\begin{figure}
	\centering
	\includegraphics[width=\linewidth]{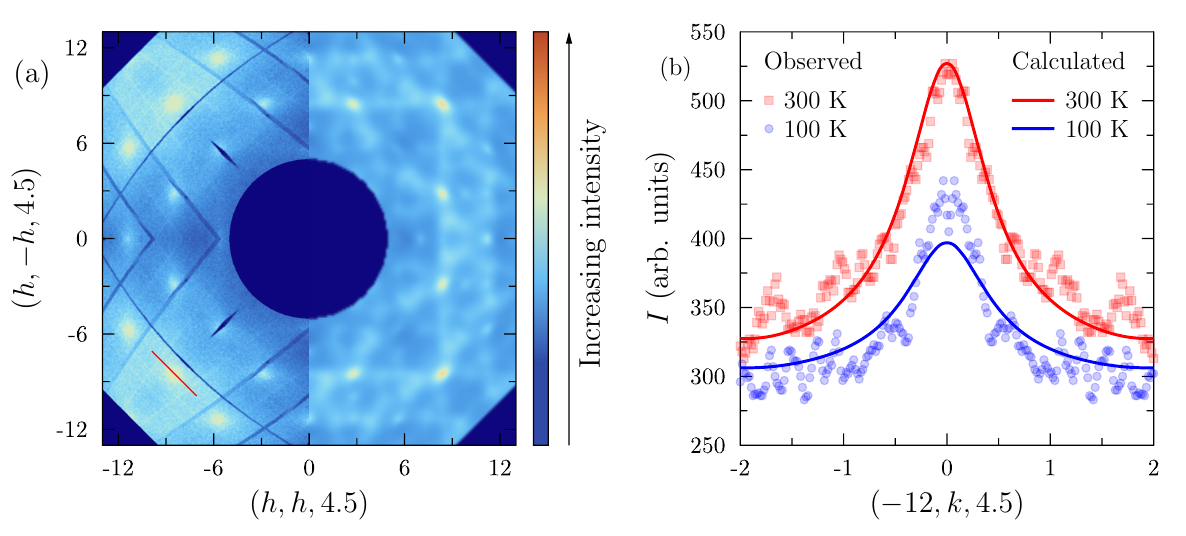}
	\caption{\textbf{The diffuse x-ray scattering from Pr$_2$Hf$_2$O$_7$}. (a, left) Experimentally measured diffuse x-ray scattering from a single crystal of Pr$_2$Hf$_2$O$_7$ in the $(h, k, 4.5)$ plane. (a, right) The phonon scattering spectra from \textit{ab initio} calculations. A cut through the $(-12, 0, 4.5)$ peak [shown as a red line in (a)] is plotted in (b), where the experimental data is compared to our calculations at both low and high temperature.}
	\label{fig:HfDiffuse}
\end{figure}

\section{Diffuse scattering from P\titlelowercase{r}$_2$H\titlelowercase{f}$_2$O$_7$}
As mentioned in the main text, we have also performed diffuse scattering measurements on another quantum spin ice candidate material, Pr$_2$Hf$_2$O$_7$.
This materials, closely related to \pzo, shares the same pyrochlore crystal structure, simply with Hf replacing Zr as the non-magnetic ion.
As such, it is reasonable to ask whether or not the disorder is the same between these two systems.
It has been reported that Pr$_2$Hf$_2$O$_7$ has a very low level of structural disorder~\cite{sibille2018experimental}.
Despite this, extensive diffuse scattering is observed in our measurements, see Fig.~\ref{fig:HfDiffuse}.

There are key differences between the diffuse scattering from Pr$_2$Hf$_2$O$_7$ and \pzo, in particular the temperature dependence.
In \pzo\ the diffuse scattering is relatively temperature-independent, supporting the origin of this scattering being defects.
Conversely, in Pr$_2$Hf$_2$O$_7$ the intensity of the diffuse scattering is significantly reduced at low temperature, suggesting a different origin.
To explore the possibility of phonons being the source of this scattering, we calculated the phonon force constant matrix from non-magnetic analogue La$_2$Zr$_2$O$_7$ with density functional theory using CASTEP~\cite{clark2005first}.
The phonon intensity from Pr$_2$Hf$_2$O$_7$ integrated over energy transfer was then calculated after mass substitution of La by Pr and Zr by Hf, and with the Bose factor correction for the given temperature using the Euphonics code~\cite{fair2022euphonic}.
The results of these calculations are shown in Fig.~\ref{fig:HfDiffuse}, where an excellent agreement is found.
To further demonstrate this agreement, we plot a representative peak (indicated on Fig.~\ref{fig:HfDiffuse}); both the peak width and temperature-dependent intensity (controlled by the Bose factor) are captured by the calculation.
Hence, we conclude that, unlike \pzo, Pr$_2$Hf$_2$O$_7$ is dominated by phonon scattering, suggesting that the level of structural disorder is much lower.
This may make Pr$_2$Hf$_2$O$_7$ the best candidate for studying quantum spin ice physics without the added complications of defect structures that we have demonstrated are prevalent in \pzo.

\bibliography{bib}